\documentclass[submitting]{nst}

\usepackage{subfigure,dcolumn}
\usepackage[T2A,T1]{fontenc}
\usepackage[russian,english]{babel}

\usepackage{listings}
\begin{document}

\title{Study of a background reconstruction method for the measurement of D-meson azimuthal angular correlations}
\thanks{This work was supported in part by the National Natural Science Foundation of China under contract Nos. 11890710, 11890714, 11905034 and the Strategic Priority Research Program of Chinese Academy of Sciences with Grant No. XDB34000000.}

\author[马龙]{L. Ma}
\email[]{malong@fudan.edu.cn}
\affiliation{Key Laboratory of Nuclear Physics and Ion-beam Application (MOE), Institute of Modern Physics, Fudan University, Shanghai 200433, China}
\author[董昕]{X. Dong}
\email[]{xdong@lbl.gov}
\affiliation{Nuclear Science Division, Lawrence Berkeley National Laboratory, Berkeley, CA 94720, USA}
\author[黄焕中]{H.Z. Huang}
\affiliation{Key Laboratory of Nuclear Physics and Ion-beam Application (MOE), Institute of Modern Physics, Fudan University, Shanghai 200433, China}
\affiliation{University of California, Los Angeles, California 90095, USA}
\author[马余刚]{Y.G. Ma}
\affiliation{Key Laboratory of Nuclear Physics and Ion-beam Application (MOE), Institute of Modern Physics, Fudan University, Shanghai 200433, China}

\begin{abstract} 

We study experimental background reconstruction methods for the measurement of $D-\overline{D}$ correlation using PYTHIA simulation. Like-Sign and Side-Band background methods that are widely used in the experimental measurements of single $D$-meson production yields were deployed for the correlation study. It is found that the Like-Sign method which well describes the combinatorial background of single $D^{0}$ meson yields fails to reproduce the correlated background in the $D^{0}-\overline{D^{0}}$ correlation measurement, while the Side-Band background method yields a good description of the background for both the single $D^0$ yields and of the correlated background of the $D^{0}-\overline{D^{0}}$ correlation measurement. We further examine the validity of the correlation methods under different signal-to-background ratios providing direct references for the experimental measurements.

\end{abstract}


\maketitle

\section{Introduction}

Quantum Chromodynamics (QCD) is the theory describing quarks, gluons and the strong interaction between them. In QCD, heavy flavor quarks ($c$, $b$) are mostly produced through initial hard scatterings in high energy collisions of nucleons or nuclei. Because of their large masses, heavy quarks may offer a unique sensitivity to study the cold and hot QCD medium created in these collisions~\cite{intro0,intro1,intro2,RR1,LHSong2018}. In proton + proton ($p+p$) collisions, perturbative QCD (pQCD) calculations reproduce the inclusive heavy flavor hadron production cross section data over a broad range of collision energies and rapidities~\cite{SPS,Tevetron1,Tevetron2,LHC1,LHC2}. The nuclear modification factor ($R_{AA}$) for charmed hadrons in heavy-ion collisions shows a significant modification compared to the $p+p$ reference~\cite{ZBTang2020}. Several models with different energy loss mechanisms can all describe the experimental data~\cite{FNOLL1,FNOLL2,PYTHIA1,intro12}. 

Recent research suggests that azimuthal correlations $\Delta\phi$ between heavy quark pairs offer new insight into charm-medium interaction dynamics, and therefore can help distinguish different energy loss mechanisms in the hot QCD medium~\cite{intro3,intro4,PBG1,PBG2,WNZhang2018}. The theoretical prediction indicates that pure radiative energy loss does not change the initial angular correlation function significantly, while pure collisional energy loss is more efficient at diluting the initial back-to-back charm pair correlation.  Furthermore, the momentum broadening in the direction perpendicular to the initial quark momentum, which cannot be probed directly with traditional single particle measurements (e.g. $R_{AA}$ and elliptic flow parameter $v_{2}$), could be reflected in the azimuthal angle correlations~\cite{intro4,intro8,SChun2020}.

In $p+p$ collisions, charm quark pairs are produced through initial back-to-back hard scatterings in leading order. In next-to-leading order, the angular correlation between charm quark pairs is widening. In particular, it will show a near-side peak at $\Delta\phi\sim0$ if the charm pairs are produced through gluon splitting. The measurement of $D-\overline{D}$ correlations in $p + p$ collisions not only provides a baseline for measurements in heavy-ion collisions, but also offers a good constraint for pQCD calculations. $D$-mesons inherit most of the initial charm pair correlations, but weak decays smear the correlation significantly. Therefore measurement of $D-\overline{D}$ correlations should be the most sensitive probe to study charm quark pair correlations~\cite{intro5,intro6,RR2,HWang2019}.

The experimental reconstruction of $D-\overline{D}$ azimuthal angular correlation is challenging. It requires reconstruction of two charmed hadrons in a single event. Charmed hadrons need to be reconstructed through their hadronic decay channels with small branching ratios. Furthermore, there are often sizable background in each reconstructed charmed hadron. In single charmed hadron yield measurements, for instance $D^{0}$ mesons through the $K^-\pi^+$ decay channel, several background methods, e.g. Like-Sign (LS), Side-Band (SB) and Mixed-Event (ME) etc. were deployed by experimentalists~\cite{intro7,intro10}. In the ME technique, background pairs are reconstructed using two daughter tracks from different events. Since the tracks are produced in different events, the background reconstructed is uncorrelated with the foreground $D^{0}$ candidates. By mixing multiple events, this method has the advantage to reproduce the combinatorial background with good statistics. In the LS technique, background is generated by pairing daughter tracks with the same charge sign. It contains the background produced correlated in pairs with opposite charge signs in the same event. In the SB technique, opposite sign pairs with invariant masses away from the $D^0$ peak are used, and usually two symmetric mass regions on both sides of the $D^0$ peak are selected and an average of these two is chosen to represent the background underneath the $D^0$ peak. To a reasonable precision, both LS and SB techniques can successfully reproduce the background in single $D^{0}$ yield measurements. 

In this paper, we investigate these background reconstruction methods for experimental measurement of $D-\overline{D}$ correlations. The ME technique misses the background correlation in the same event and it typically needs to be normalized to either LS or SB distributions. In the following study we focus on the comparison of the LS and SB background techniques.

\section{PYTHIA study for $D-\overline{D}$ correlations}

The Monte Carlo event generator PYTHIA (version 8.168) is used for this study~\cite{tune0}. We focus on $p+p$ collisions at $\sqrt{s}$ = 500 GeV for illustration. The parameters were tuned so that PYTHIA can reproduce the experimental data on inclusive $c\bar{c}$ production cross-section in $p+p$ collisions at 500 GeV measured by the STAR experiment at RHIC~\cite{tune1}.   

\begin{figure}[htbp]
\includegraphics[width=0.45\textwidth,height=0.4\textwidth]{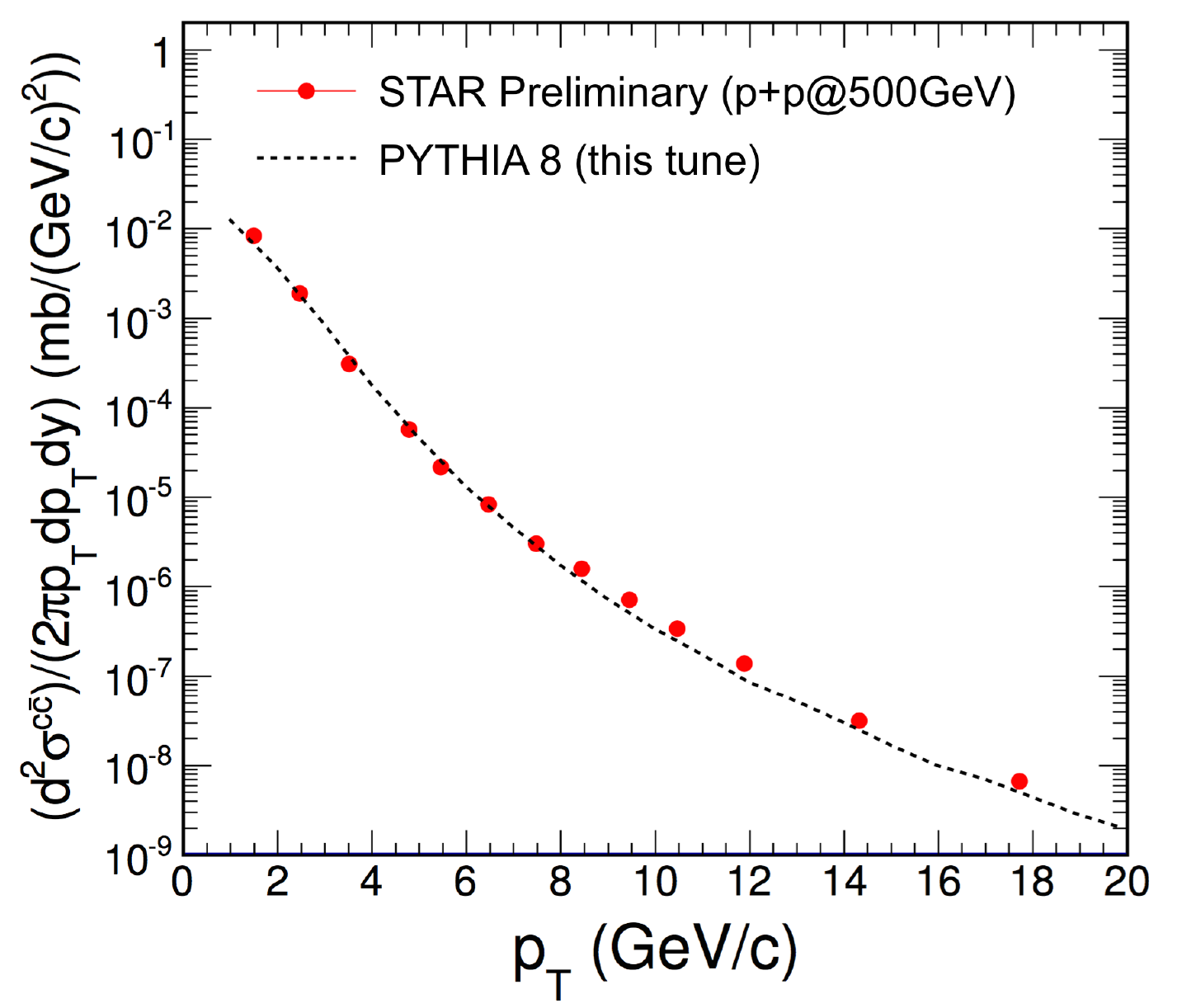}
\caption{(Color online) Charm pair cross section as a function of transverse momentum in p + p collisions at $\sqrt{s}$ = 500 GeV in PYTHIA (dashed line) compared with STAR measurements (solid circles).}
\label{crosssection}
\end{figure}

Fig.~\ref{crosssection} shows the $c\bar{c}$ production cross section as a function of transverse momentum in PYTHIA in comparison with the STAR measurements. The modified PYTHIA parameters in this tune are: strong interaction coupling constant ($\alpha_{s}$) value of final parton shower (TimeShower:alphaSvalue) to be 0.15 and minimum invariant transverse momentum ($p_{T}$) threshold for hard QCD process (PhaseSpace:pTHatMin) to be 1.5 GeV/c. PYTHIA with this tune describes both the magnitude and the $p_{T}$ spectrum well. It was also found that changing these two parameters has a negligible effect on charm correlations.

A sample of six billion PYTHIA minimum bias events with this tune were generated for the $D^{0}-\overline{D^{0}}$ correlation study. $D^{0}$ mesons can be directly accessed in the PYTHIA simulation based on their particle identification number. To emulate the experimental measurement, $D^{0}$s were reconstructed by pairing kaon and pion candidate pairs via the typical hadronic decay channel $D^{0} \rightarrow K^{-} + \pi^{+}$ and its charge conjugate channel for $\overline{D^{0}}$. In a real experiment with a silicon vertex detector, many background tracks from the primary collisions can be eliminated, but there remains considerable background, particularly in the low $p_{T}$ region. 

In this study, we don't distinguish secondary decay vertices in the $D^0$ reconstruction. Instead, we combine all kaons and pions at midrapidity ($|\eta| \leq 1$) in the final stage in the PYTHIA output, which allows us to study the validity of the background reconstruction methods with different signal-to-background ratios of the reconstructed $D^0$ candidates. The invariant mass of Unlike-Sign and Like-Sign kaon and pion pairs in the same event are calculated. A finite momentum resolution effect is included so that the reconstructed $D^0$ signal peak has the width observed by the experiment.

\begin{figure}[htbp]
\includegraphics[width=0.45\textwidth,height=0.4\textwidth]{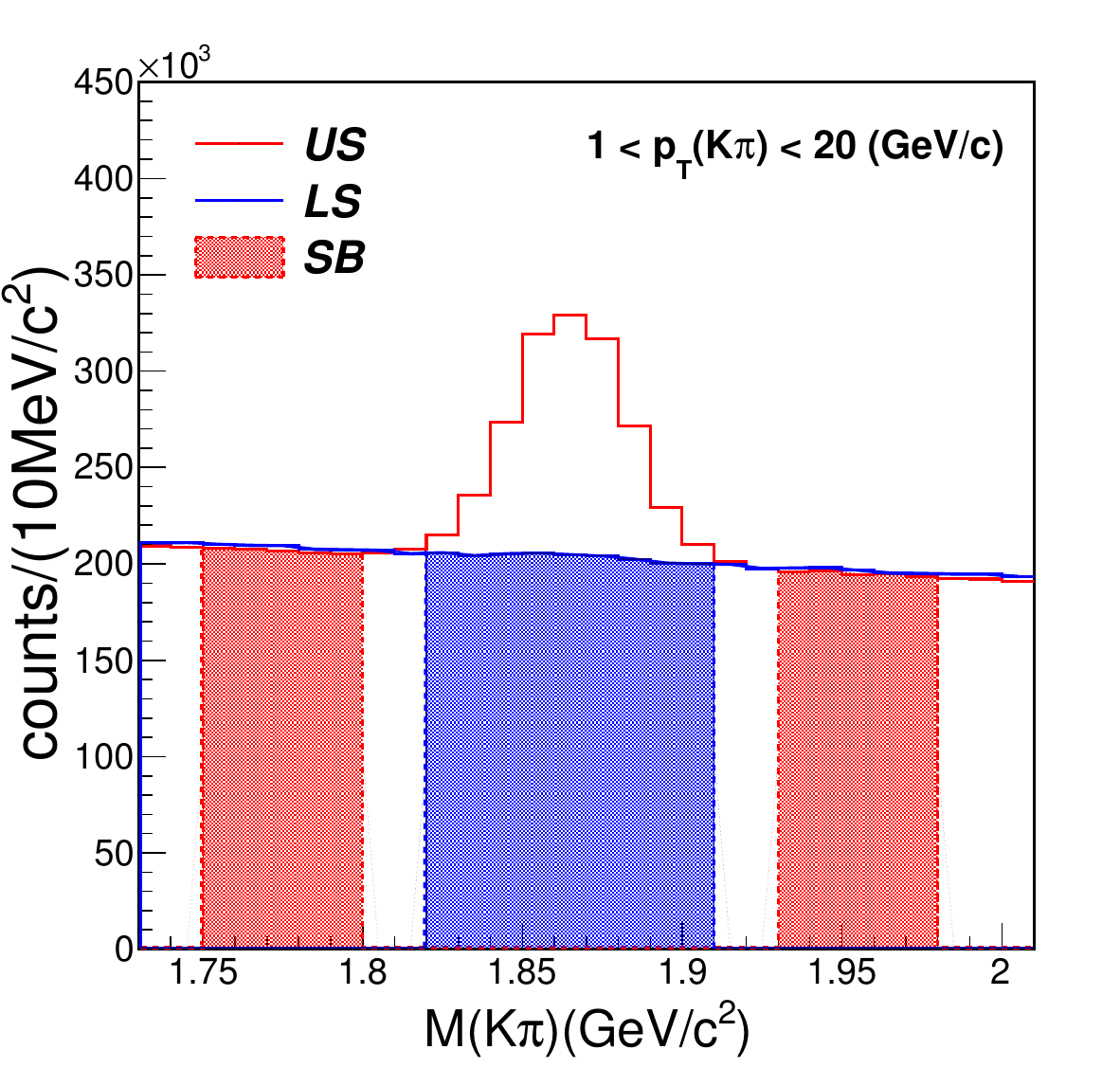}
\caption{(Color online) Invariant mass distribution of all final stage kaon and pion pairs with opposite signs in PYTHIA data at mid-rapidity (shown by solid red line, US). Like-Sign method reproduces the combinatorial background shown by the blue solid line (LS). The blue shaded region shows the Like-Sign background within a $\pm$3$\sigma$ window of the signal peak. The Side-Band background regions are shaded red (SB).}
\label{invmass}
\end{figure}

Fig.~\ref{invmass} demonstrates the $D^{0}$($\overline{D^{0}}$) signal and the combinatorial background from the Like-Sign method and the Side-Band method. The Like-Sign background and Side-Band background regions are denoted by the blue and red hatched areas, respectively. Invariant mass distribution of $\overline{D^{0}}$ is almost identical to $D^{0}$ in both shape and magnitude. The background is found to be flat in PYTHIA within a relative wide invariant mass range. For simplicity, we denote $D^{0}$($\overline{D^{0}}$) candidates from $K^{-}$$\pi^{+}$($K^{+}$$\pi^{-}$) pairs with unlike signs as `US' candidates, and those from $K^{-}$$\pi^{-}$ or $K^{+}$$\pi^{+}$ pairs with same charge sign as `LS' background. The Side-Band background is denoted as `SB'. Fig.~\ref{invmass} shows both LS and SB methods can reasonably reproduce the real background underneath the reconstructed $D^0$ signals. For the single particle yield measurement, the $D^0$ and $\overline{D^0}$ counts are calculated by Eq.~\ref{q00} and Eq.~\ref{q01} for LS and SB background methods, respectively.

If the background methods work well for the $D^{0}-\overline{D^0}$ correlation measurement, the correlation signal between $D^0$ and $\overline{D^0}$ can be derived using Eq.~\ref{q1} and Eq.~\ref{q2}. The asterisks (*) indicate the correlation functions between the pairs. We can also derive the $D^{0}-\overline{D^0}$ correlation signal from the PYTHIA simulation directly and compare to the reconstructed signals with these two background methods.

\begin{equation}
\begin{aligned}
N^{D+\overline{D}}_{LS} = US(K^-\pi^+) + US(K^+\pi^-)\\
- LS(K^-\pi^-) - LS(K^+\pi^+)
\end{aligned}
\label{q00}
\end{equation}

\begin{equation}
\begin{aligned}
N^{D+\overline{D}}_{SB} = US(K^-\pi^+) + US(K^+\pi^-)\\
- SB(K^-\pi^+) - SB(K^+\pi^-)
\end{aligned}
\label{q01}
\end{equation}

\begin{equation}
\begin{aligned}
C^{D\overline{D}}_{LS} = US(K^{-}\pi^{+})*US(K^{+}\pi^{-})\\
-LS(K^{-}\pi^{-})*US(K^{-}\pi^{+})\\
-LS(K^{+}\pi^{+})*US(K^{+}\pi^{-})\\
+LS(K^{-}\pi^{-})*LS(K^{+}\pi^{+}),
\end{aligned}
\label{q1}
\end{equation}

\begin{equation}
\begin{aligned}
C^{D\overline{D}}_{SB} = US(K^{-}\pi^{+})*US(K^{+}\pi^{-})\\
-SB(K^{+}\pi^{-})*US(K^{-}\pi^{+})\\
-SB(K^{-}\pi^{+})*US(K^{+}\pi^{-})\\
+SB(K^{-}\pi^{+})*SB(K^{+}\pi^{-}),
\end{aligned}
\label{q2}
\end{equation}

\begin{figure}[htbp]
\centering
\includegraphics[width=0.45\textwidth,height=0.38\textwidth]{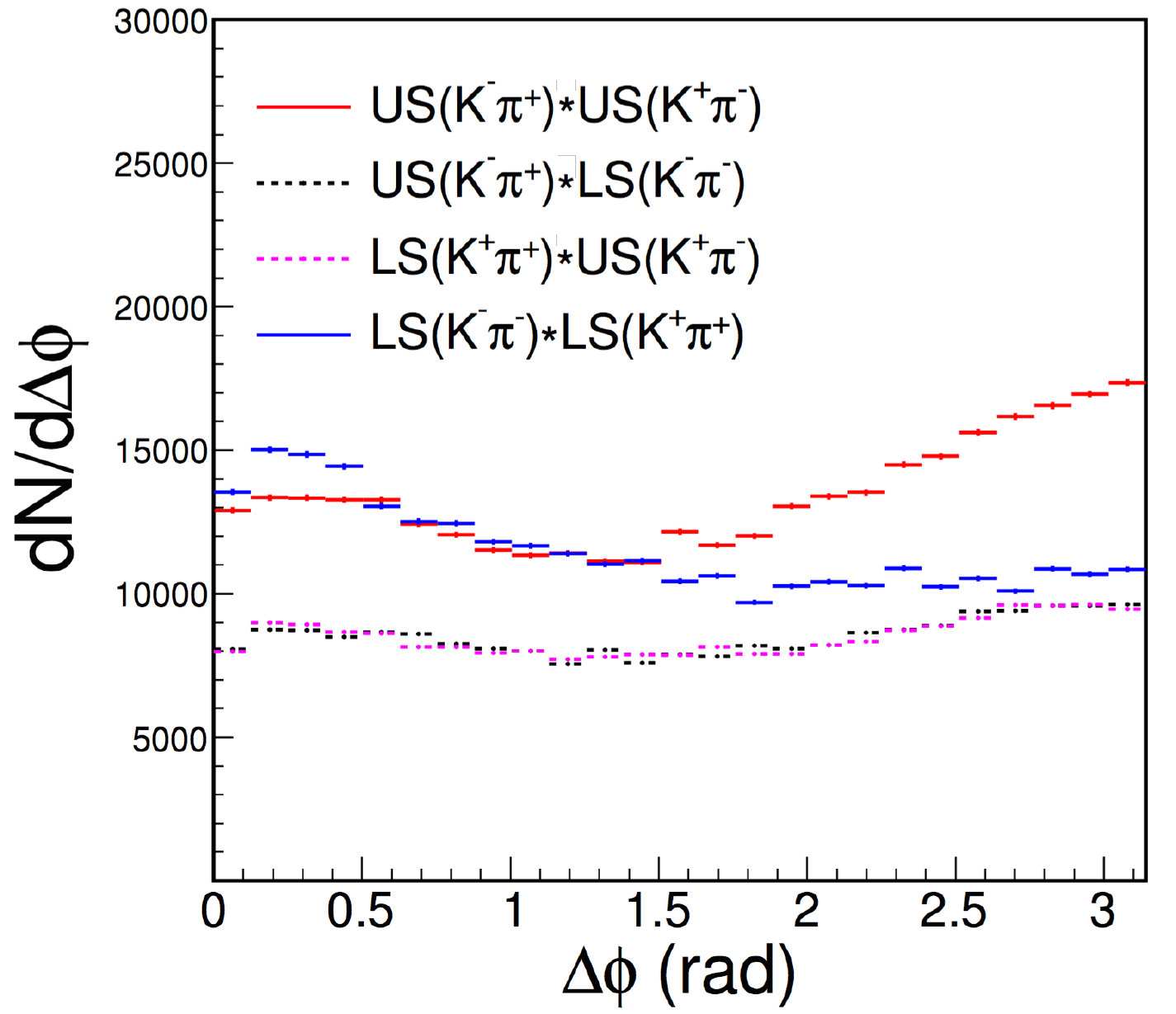}
\vspace{0.2\textwidth}%
\includegraphics[width=0.45\textwidth,height=0.38\textwidth]{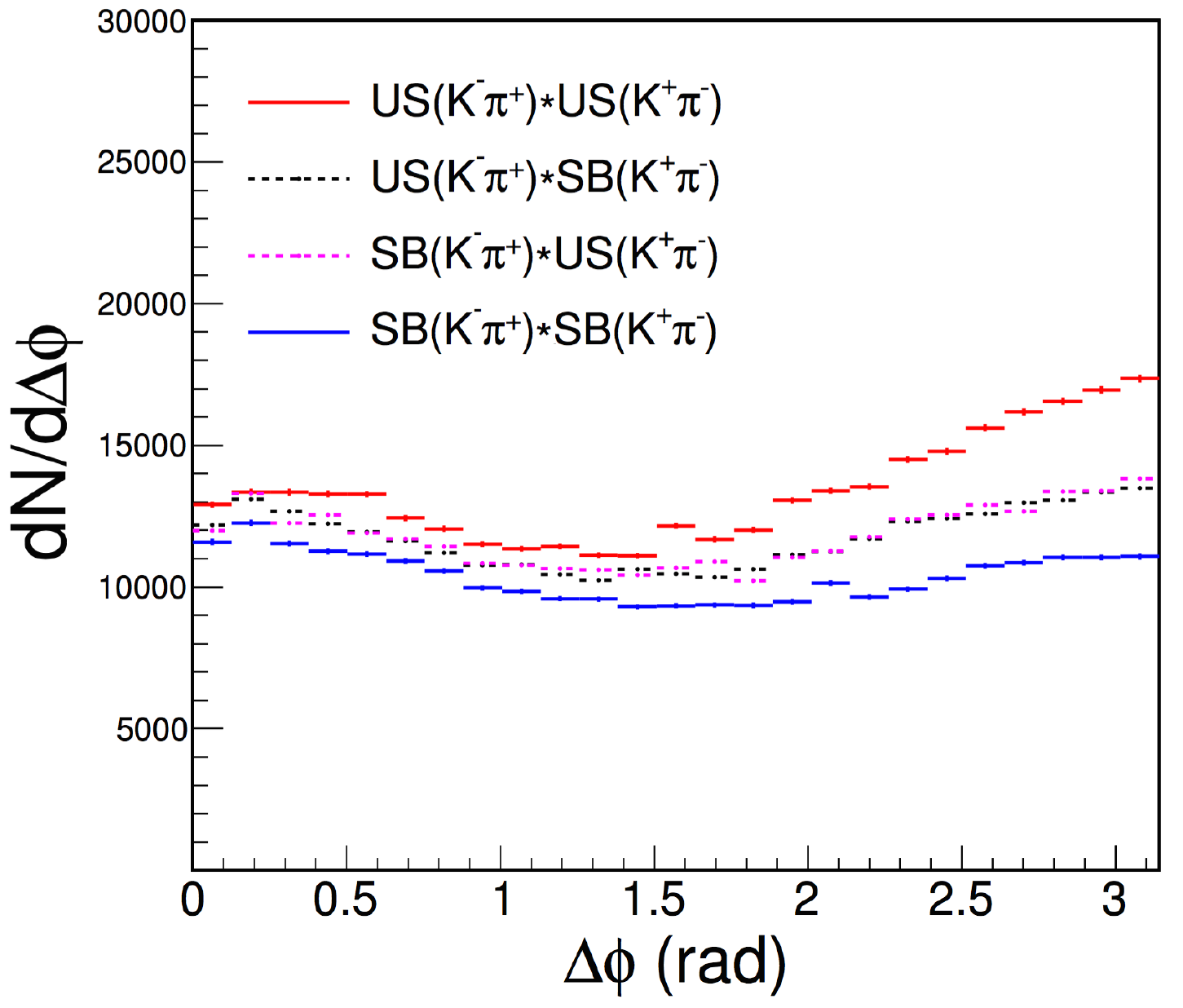}
\caption{(Color online) Upper Panel: cross correlations of $D^{0}-\overline{D^{0}}$ from Unlike-Sign candidates(US) and Like-Sign backgrounds(LS). The trigger and associate $p_{T}$ cuts are both set to 1 GeV/c with a signal-to-background ratio of around 0.3. Lower Panel: similar results from the Side-Band(SB) method.}
\label{lssbQA1} 
\end{figure}

The di-hadron correlation measurements are usually plotted as a function of the azimuthal angle difference, i.e. $\Delta\phi = \phi_{D^0} - \phi_{\overline{D^0}}$. Fig.~\ref{lssbQA1} upper panel shows the correlations between Unlike-Sign candidates and the Like-Sign backgrounds as a function of $\Delta\phi$.  The $p_{T} > 1.0$ GeV/c cut is set for both $D^0$ and $\overline{D^0}$ mesons, and the mass window cuts for US, LS and SB pairs used are shown as the colored bands in Fig.~\ref{invmass}.  The plot shows that the correlation between Like-Sign and Like-Sign background pairs (LS*LS) tends to peak at $\Delta\phi$ around 0 and that its magnitude is considerably larger than that between Like-Sign background and UnLike-Sign candidates (LS*US).  The lower panel in Fig.~\ref{lssbQA1} shows the results of the Side-Band method with the same trigger $p_{T}$ and S/B ratio as the Like-Sign method. The correlation between the Side-Band background and Unlike-Sign candidates lies between the other two correlation terms. The correlation between two Side-Band background pairs shows a similar trend as that between Side-Band background and Unlike-Sign candidates. 

\begin{figure}[htbp]
\centering
\includegraphics[width=0.45\textwidth,height=0.4\textwidth]{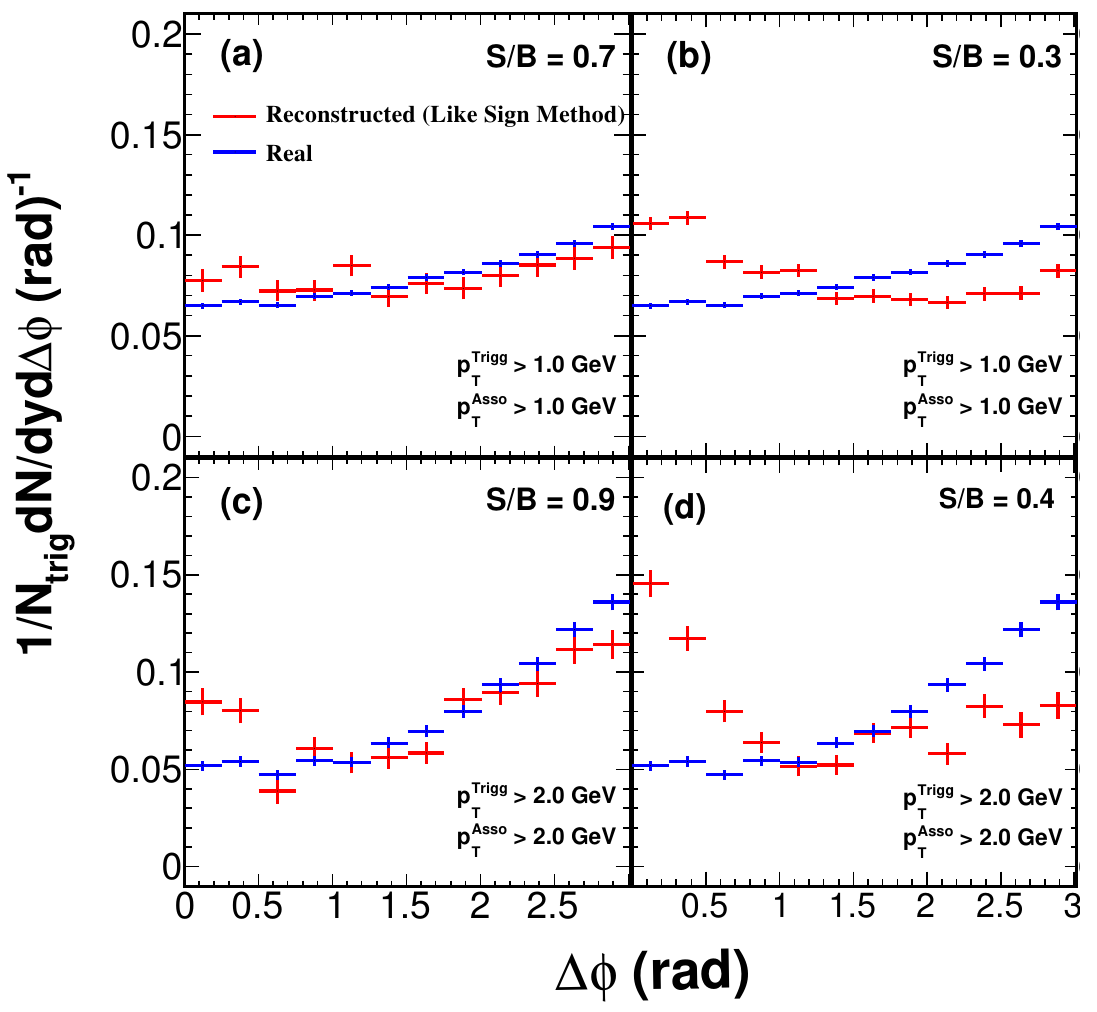}
\vspace{0.2\textwidth}%
\includegraphics[width=0.45\textwidth,height=0.4\textwidth]{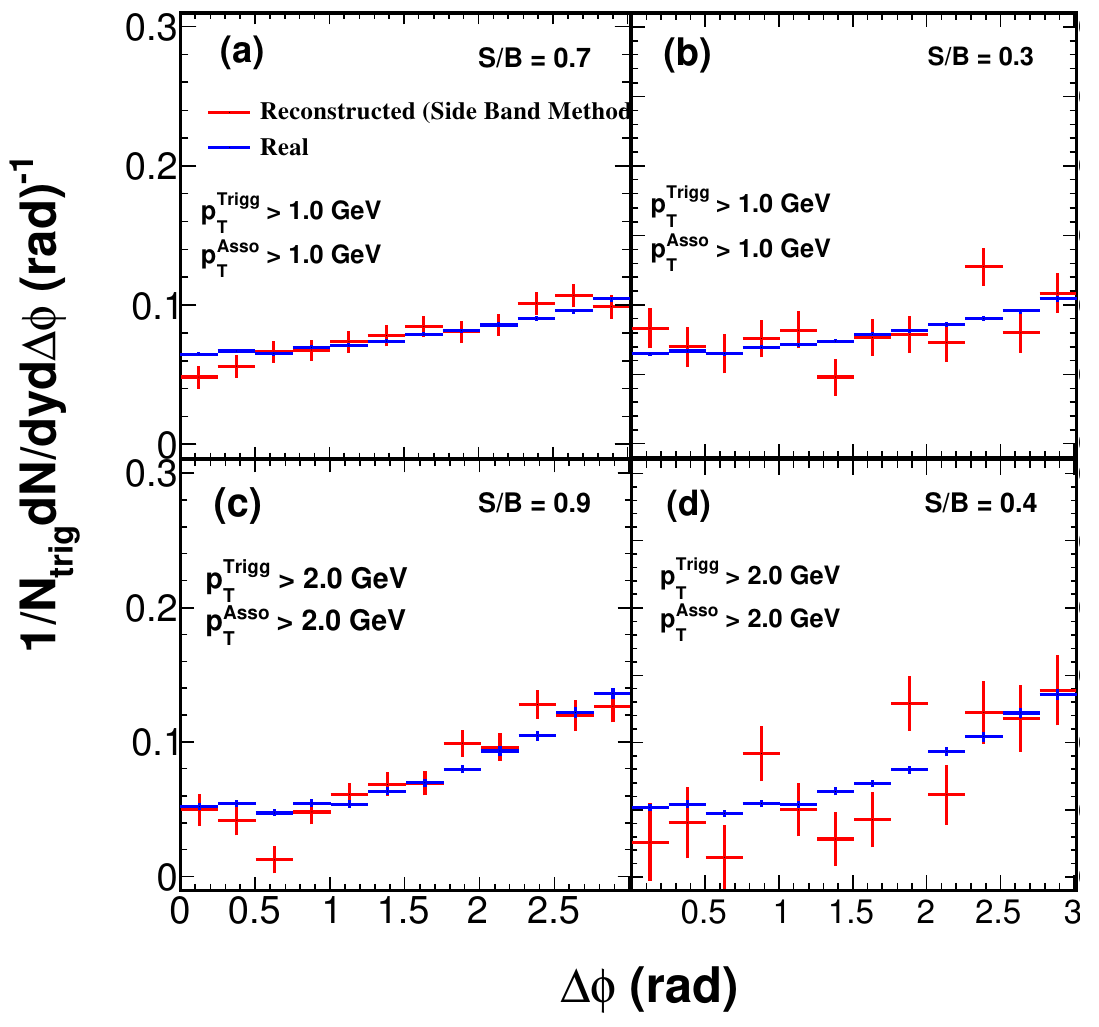}
\caption{(color online). The $D^{0}-\overline{D^{0}}$ correlation as a function of relative azimuthal angle $\Delta\phi$ in $p+p$ collisions at $\sqrt{s}$ = 500 GeV calculated with Like-Sign method (Upper Panel) and Side-Band method (Lower Panel) based on Eq.~\ref{q1} in PYTHIA simulation. The transverse momentum dependence is shown with $p_{T}$ cuts applied to triggered and associated $D$ mesons. Panels (a)-(d) show correlations of reconstructed $D^{0}$ mesons under different S/B ratios in comparison with correlations of produced $D^{0}-\overline{D^{0}}$ pairs in PYTHIA.}
\label{lssbQA2}
\end{figure}

In Fig.~\ref{lssbQA2}, the reconstructed $D^0-\overline{D^0}$ correlation signals with LS and SB background methods are compared to the real correlation signals from PYTHIA directly. Two sets of $p_{T}$ cuts were imposed for triggered and associated $D$-mesons, shown in the upper and lower panels respectively. Panels in two different columns show the comparisons with two different mass window cuts which result in different signal-to-background ratios of the reconstructed $D^0$ candidates. The red data points represent the correlation signals from reconstructed $D^{0}$s, and the blue data points are the real $D^{0}$ correlations directly obtained from PYTHIA with the same kinematic cuts applied. Similar results from the SB method are also shown in Fig.~\ref{lssbQA2}. 

As we can see, reconstructed correlations using the LS method are different from the real $D^{0}-\overline{D^0}$ signal from PYTHIA. Particularly, the reconstruction correlations start to show an enhanced structure in the near side region when the signal-to-background ratio is getting smaller. While reconstructed correlations using the SB method can reproduce the real $D^{0}-\overline{D^0}$ signal reasonably well in these kinematic and signal-to-background ratio regions.  It is also found that the quality of the agreement does not depend on the transverse momentum cut. It depends on the signal-to-background ratios of the $D^0$ candidates.

\begin{equation}
u_{i}=(\frac{1}{N_{trig}}\frac{dN}{d\Delta\phi_{i}})_{reco},\qquad
v_{i}=(\frac{1}{N_{trig}}\frac{dN}{d\Delta\phi_{i}})_{real}
\label{q3}
\end{equation}

\begin{equation}
\Delta P=\frac{1}{N}\sum_{i=1}^{k} \left | \frac{u_{i}-v_{i}}{v_{i}} \right |,\qquad
\Delta E=\frac{1}{N}\sum_{i=1}^{k} \left | \frac{\sqrt{\sigma^{2}_{u_{i}}+\sigma^{2}_{v_{i}}}}{v_{i}} \right |
\label{q4}
\end{equation}

\begin{figure}[htbp]
\centering
\includegraphics[width=0.4\textwidth,height=0.35\textwidth]{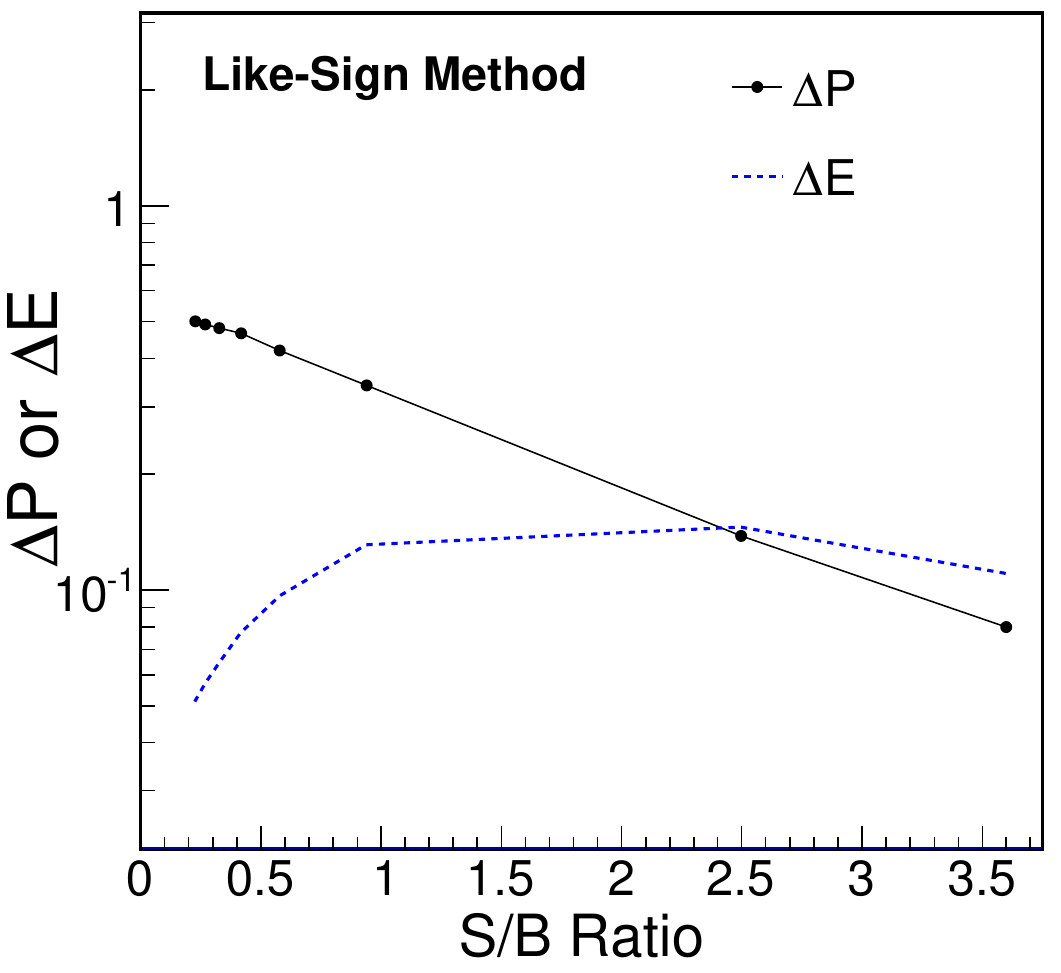}
\vspace{0.2\textwidth}%
\includegraphics[width=0.4\textwidth,height=0.35\textwidth]{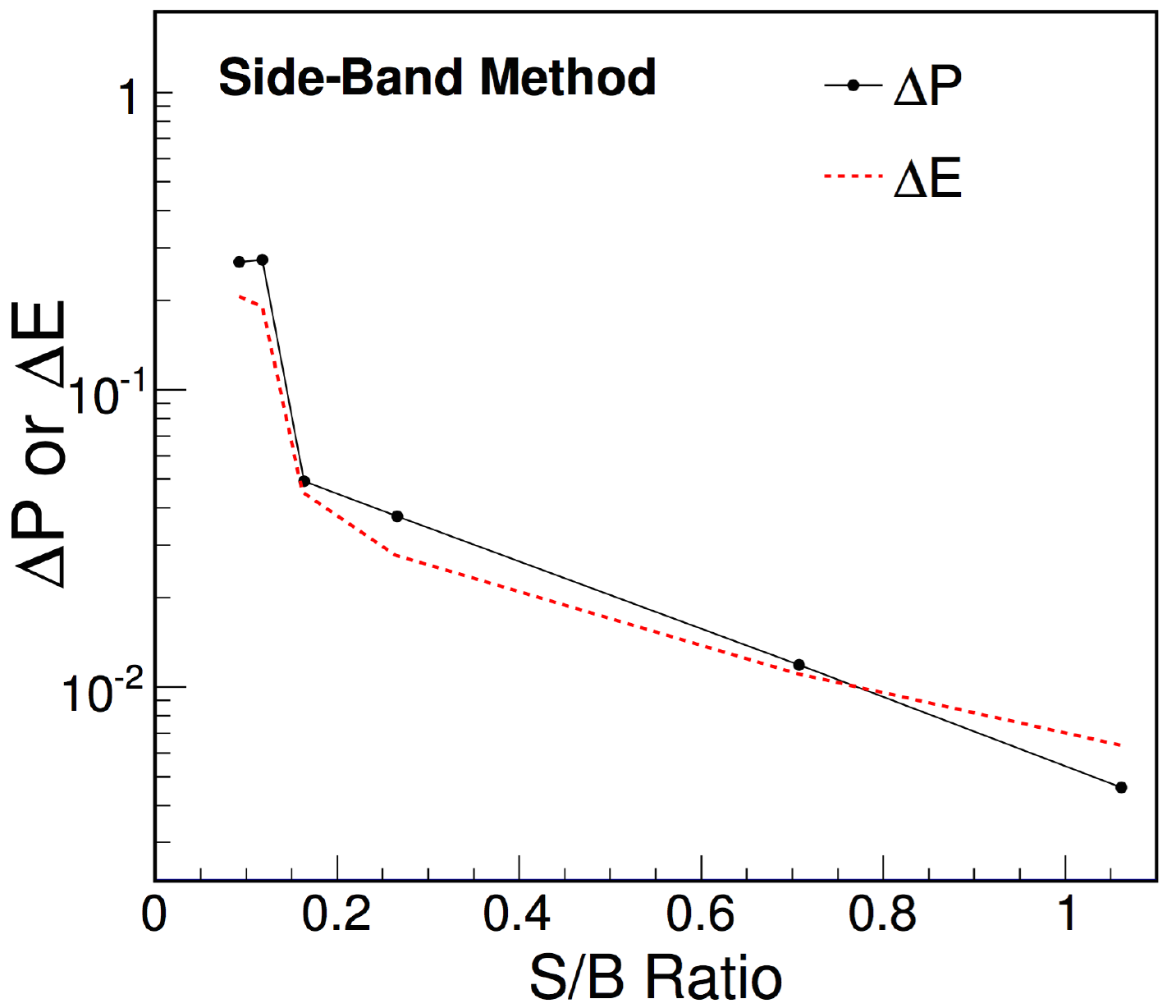}
\caption{(color online). The summary plots of the goodness of fit calculated with Like-Sign method (Upper Panel) and Side-Band method (Lower Panel) in PYTHIA simulation. The estimator is shown as a function of the signal over background ratio (S/B). Solid line and dashed line show $\Delta$P and $\Delta$E respectively.}
\label{lssb-sum} 
\end{figure}

To better illustrate the performance of these two background methods in measuring the angular correlations of the $D-\overline{D}$ pairs, we define two variables to quantify the goodness of fit for the reconstructed correlation signals with respect to the real $D^{0}-\overline{D^{0}}$ correlations from PYTHIA. $\Delta P$ and $\Delta E$ are defined in Eq.~\ref{q4} to describe the relative difference of the data points and the statistical errors from this sample. $u_{i}$ and $v_{i}$ are values of the number $i$ data points of reconstructed and real correlation signals in each $\Delta\phi$ bin.  $N$ is the total number of data points in each correlation signal assuming the same binning for the histograms. Fig.~\ref{lssb-sum} shows the corresponding results from Like-Sign method and Side-Band methods respectively. As we can see, $\Delta P$ in Like-Sign results shows a large deviation from $\Delta E$ when the S/B ratio goes down indicating that Like-Sign method fails to reproduced real correlation at relatively low S/B ratios. While the Side-Band method shows a good performance throughout the whole S/B ratio region investigated. The increase of both $\Delta P$ and $\Delta E$ at low S/B region in the Side-Band method is due to the reduced statistics. We also study the performance of the two background methods with considering $D^{0}$ from $D^{*}$ decay and non-prompt $D^{0}$ from B-decay. It is found the conclusions of the goodness of fit for both methods stay unchanged. Experimentally, as particle mis-identification (Mis-PID) may place effect on the background reconstruction and cause double-counting of the signals, we further evaluate such effects on the correlation reconstruction through a toy Monter Carlo simulation based on the PID criteria for $p+p$ collisions in STAR analysis~\cite{tune1}. We found the mis-PID effect is quite tiny (<1$\%$) in this case.

\begin{figure}[htbp]
\centering
\includegraphics[width=0.4\textwidth,height=0.35\textwidth]{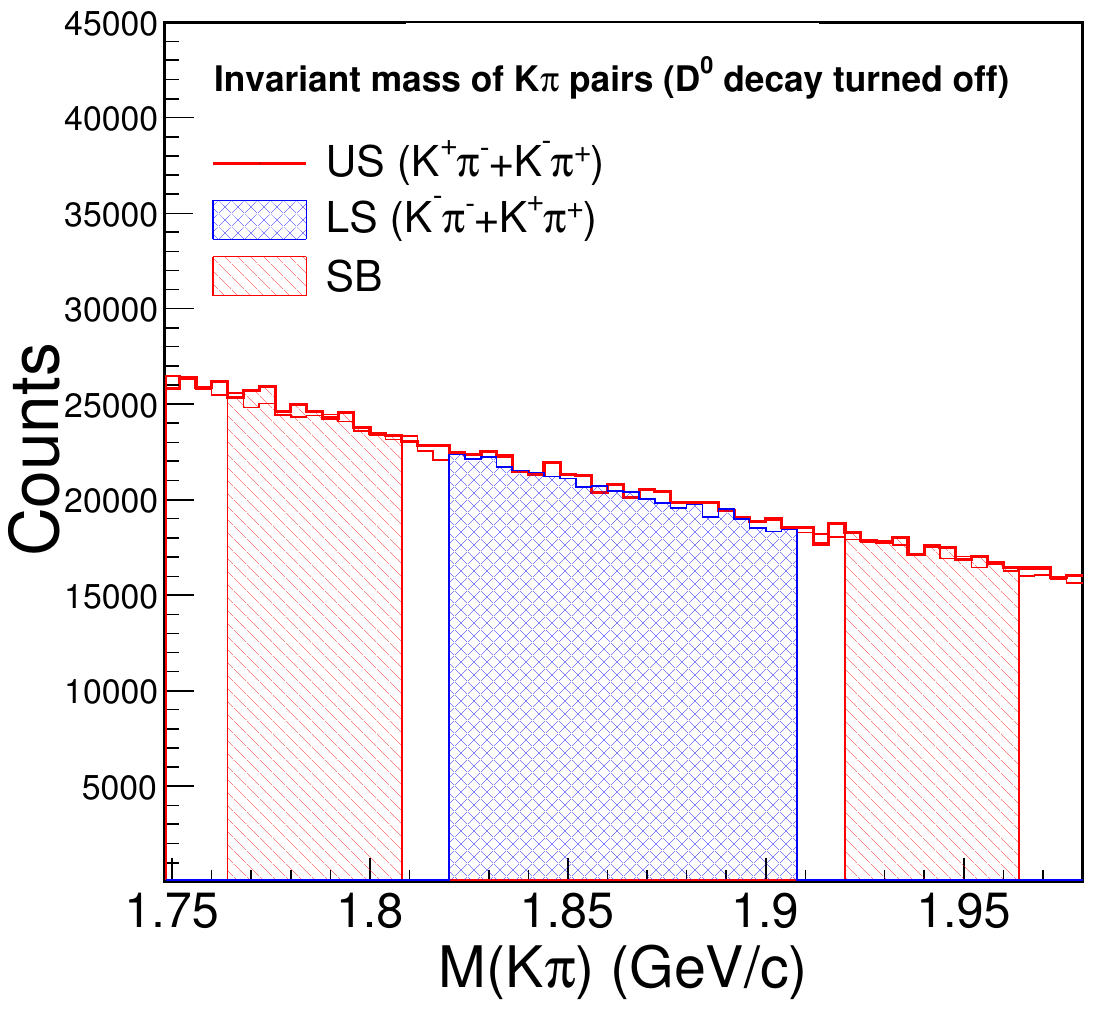}
\caption{(Color online)Invariant mass distribution of pure $K\pi$ pairs in PYTHIA. $D^{0}->K\pi$ hadronic decay process is turned off. Red line: Unlike-Sign(US) $K\pi$ pairs. Blue hatched area: Like-sign(LS) $K\pi$ pairs within cut window. Red shaded area: Side-Band(SB) $K\pi$ pairs within cut window.}
\label{invmass_nocc}
\end{figure}

Considering that in the Like-Sign method, when a $K^+\pi^+$ pair is selected, there is a higher probability for finding a $K^-\pi^-$ pair than another $K^+\pi^+$ pair due to local and global charge conservation. The reconstructed correlation signal after Like-Sign background subtraction from Eq.~\ref{q1} should contain all correlations between $K^+\pi^-$ and $K^-\pi^-$ pairs including the $D^0-\overline{D^0}$ correlation we are interested in as well as the correlation due to charge conservation. To further demonstrate that the additional correlation observed in the Like-Sign method is related to the underlying event instead of the $D^0-\overline{D^0}$ signal, we turn off $D^{0}$ hadronic decay process in the PYTHIA simulation and run the same analysis.

\begin{figure}[htbp]
\centering
\includegraphics[width=0.4\textwidth,height=0.55\textwidth]{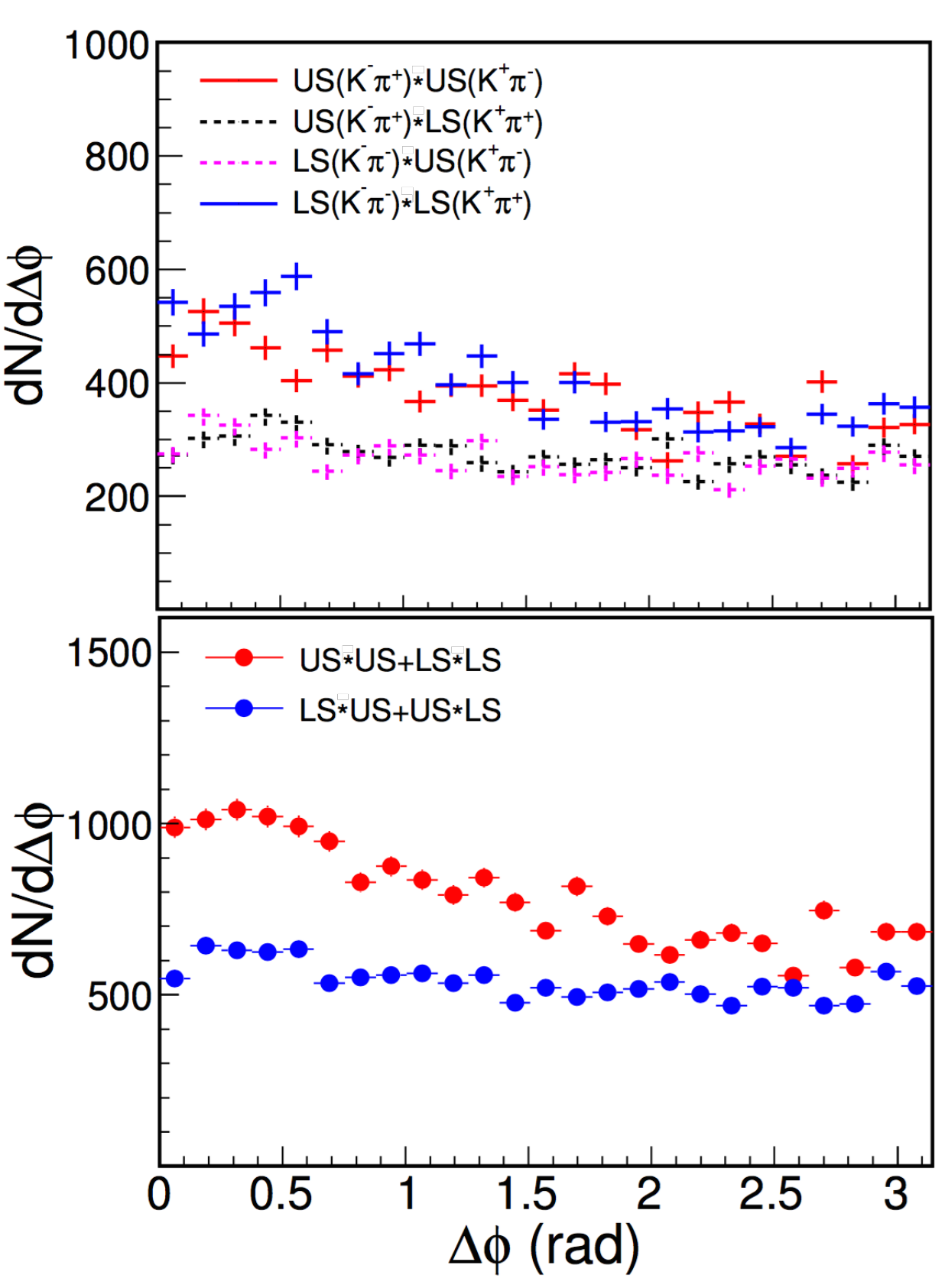}
\caption{(Color online) Cross correlations of the pure Like-Sign(LS) and Unlike-Sign(US) $K\pi$ pairs in Like-Sign method.}
\label{noccQA1} 
\end{figure}

\begin{figure}[htbp]
\centering
\includegraphics[width=0.4\textwidth,height=0.55\textwidth]{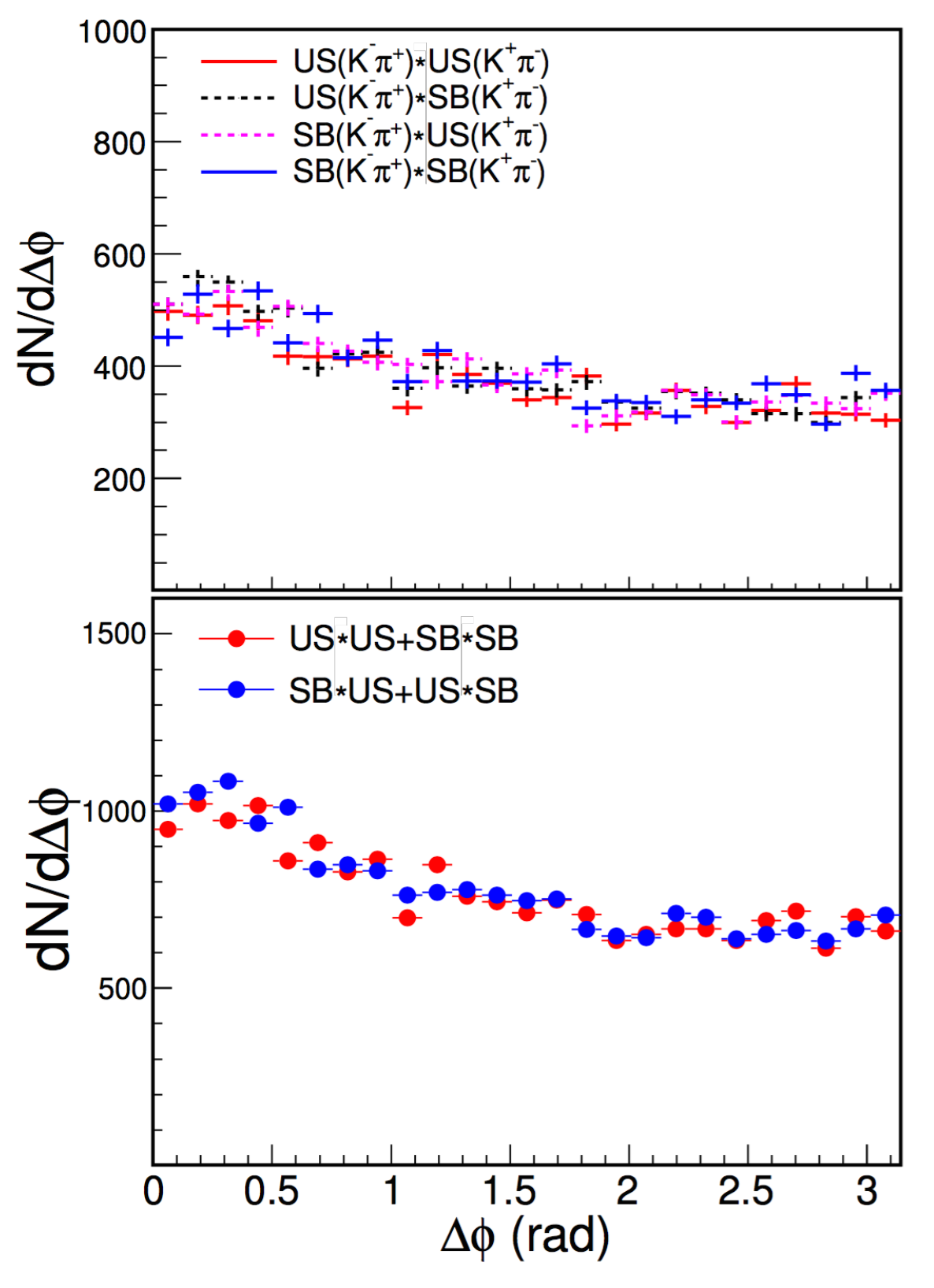}
\caption{(Color online) Cross correlations of Side-Band(SB) and Unlike-Sign(US) $K\pi$ pairs in Side-Band method.}
\label{noccQA2} 
\end{figure}

Fig.~\ref{invmass_nocc} shows the invariant mass distribution of pure $K\pi$ pairs without $D^{0}$ decay contribution. Cross correlations between Unlike-Sign/Like-Sign, Like-Sign/Like-Sign pairs are plotted in comparison with the Unlike-Sign/Unlike-Sign pair correlations in Fig.~\ref{noccQA1} with different cuts applied to the invariant mass region. Similarly, results from the Side-Band method are shown in Fig.~\ref{noccQA2}. There is a large difference between the LS*US pair correlation and the US*US pair correlation while there is very small difference between LS*LS and US*US. This is consistent with our understanding that there is an additional correlation not originating from the $D^0-\overline{D^0}$ pairs.
  
The Side-Band method is not affected by charge conservation. It is shown that all cross correlations fall in the same trend, and there is no remaining $K^+\pi^-$-$K^-\pi+$ correlation when the $D^0\rightarrow K^+\pi^-$ decay is turned off.

\section{Conclusion}

In summary, we study the background reconstruction methods for azimuthal correlations between $D^{0}$ and $\overline{D^{0}}$ pairs by a PYTHIA simulation. 

Both Like-Sign and Side-Band methods show a good description of the background when reconstructing single $D^0$ yields. However, when reconstructing the correlation signal, the Like-Sign method fails to reproduce the $D^{0}-\overline{D^0}$ correlation. The residual correlation after Like-Sign background subtraction mainly comes from the underlying event activity likely due to local or global charge conservation. We demonstrate that the Side-Band method shows a good performance in describing the correlation background and therefore reproduce the original $D^{0}-\overline{D^0}$ correlation in the signal-to-background rate regions investigated. The upcoming sPHENIX experiment at RHIC will explore the charm correlation in $p+p$ and Au+Au collisions through measuring the $D-\overline{D}$ azimuthal correlation with fully reconstruction of D-mesons through their hadronic decay channels. Our study of the correlation methods provides important references for the future experimental measurements.

\end{document}